\documentclass[
backend=bibtex,
reprint,
superscriptaddress,
preprintnumbers,
nobibnotes,nofootinbib,
amsmath,amssymb,
aps,
floatfix,
]{revtex4-1}

\usepackage{gensymb}

 \usepackage[center]{titlesec}
  \titleformat{\section}
   {\centering\normalfont\fontsize{10}{10}\bfseries}{\thesection}{1em}{}
  \titlespacing{\section}{0pt}{12pt plus 4pt minus 2pt}{6pt plus 2pt minus 2pt}

\usepackage{textcomp}
\usepackage{cancel}
\usepackage{graphicx}
\usepackage{natbib}
\usepackage{mathrsfs}
\usepackage{ulem}
\usepackage{siunitx}
\usepackage{textgreek}
\usepackage{gensymb}
\usepackage{color}
\usepackage{upgreek}
\usepackage[linktoc=page,colorlinks,urlcolor=blue,citecolor=blue,linkcolor=blue]{hyperref}

\newcommand{\unm}{Center for High Technology Materials and Department of Physics and Astronomy, University of New Mexico, Albuquerque, 87106 NM, USA}

\begin{document}
\title{Two-dimensional nuclear magnetic resonance spectroscopy with a microfluidic diamond quantum sensor}

\date{\today}
\author{Janis Smits$^{\mathsection}$}
\affiliation{\unm}
\affiliation{Laser Center of the University of Latvia, Riga, LV-1586, Latvia}

\author{Joshua Damron$^{\mathsection}$}
\affiliation{\unm}

\author{Pauli Kehayias}
\affiliation{\unm}
\affiliation{Sandia National Laboratories, Albuquerque, 87123 NM, USA}

\author{Andrew F. McDowell}
\affiliation{NuevoMR LLC, Albuequerque, 87106 NM, USA}

\author{Nazanin Mosavian}
\affiliation{\unm}

\author{Ilja Fescenko}
\affiliation{\unm}

\author{Nathaniel Ristoff}
\affiliation{\unm}

\author{Abdelghani Laraoui}
\affiliation{\unm}

\author{Andrey Jarmola}
\email{andrey.jarmola@odmrtechnologies.com}
\affiliation{Department of Physics, University of California, Berkeley, 94720 CA, USA}
\affiliation{ODMR Technologies Inc., El Cerrito, 94530 CA, USA}

\author{Victor M. Acosta}
\email{vmacosta@unm.edu}
\affiliation{\unm}

\renewcommand{\thefootnote}{}{\footnote{$\mathsection$ J. Smits and J. Damron contributed equally to this study.}}

\date{\today}
\begin{abstract}
Quantum sensors based on nitrogen-vacancy centers in diamond have emerged as a promising detection modality for nuclear magnetic resonance (NMR) spectroscopy owing to their micron-scale detection volume and non-inductive based detection. A remaining challenge is to realize sufficiently high spectral resolution and concentration sensitivity for multidimensional NMR analysis of picoliter sample volumes. Here, we address this challenge by spatially separating the polarization and detection phases of the experiment in a microfluidic platform. We realize a spectral resolution of $0.65\pm0.05~{\rm Hz}$, an order-of-magnitude improvement over previous diamond NMR studies. We use the platform to perform two-dimensional correlation spectroscopy of liquid analytes within an effective ${\sim}20$ picoliter detection volume. The use of diamond quantum sensors as in-line microfluidic NMR detectors is a significant step towards applications in mass-limited chemical analysis and single cell biology. \end{abstract}

\maketitle

\section{Introduction}
Nuclear magnetic resonance (NMR) spectroscopy is a powerful and well-established method for compositional, structural, and functional analysis used in a wide range of scientific disciplines. Conventional NMR spectrometers rely on the inductive detection of oscillating magnetic fields generated by precessing nuclear spins. The signal-to-noise ratio (SNR) is strongly dependent on the external field strength ($B_0$), scaling $\propto B_0^{7/4}$ \cite{fukushimaexperimental}. This has motivated the development of increasingly large and expensive superconducting magnets to improve SNR, resulting in a twofold increase in field strength in the last $25$ years \cite{ardenkjaer2015facing}. However, even for fields exceeding $10~{\rm T}$, detection of microscale volumes often requires isotopic labeling, concentrated samples, and long experimental times \cite{ardenkjaer2015facing}

To improve sensitivity for small volume samples, miniature inductive coils have been developed \cite{stocker1997nanoliter, MCD_200774}. This approach has enabled several advances including the spectroscopy of individual egg cells \cite{fugariu2017towards, grisi2017nmr} and \textit{in vitro} diagnostics based on NMR relaxometry \cite{demas2011magnetic}. However, the present sensitivity and detection volumes are suboptimal for metabolic analysis of single mammalian cells \cite{ZEN2013} or incorporation into in-line microfluidic assays \cite{BRK2011}.
 
\begin{figure*}[ht]
    \centering
    \includegraphics[width=\textwidth]{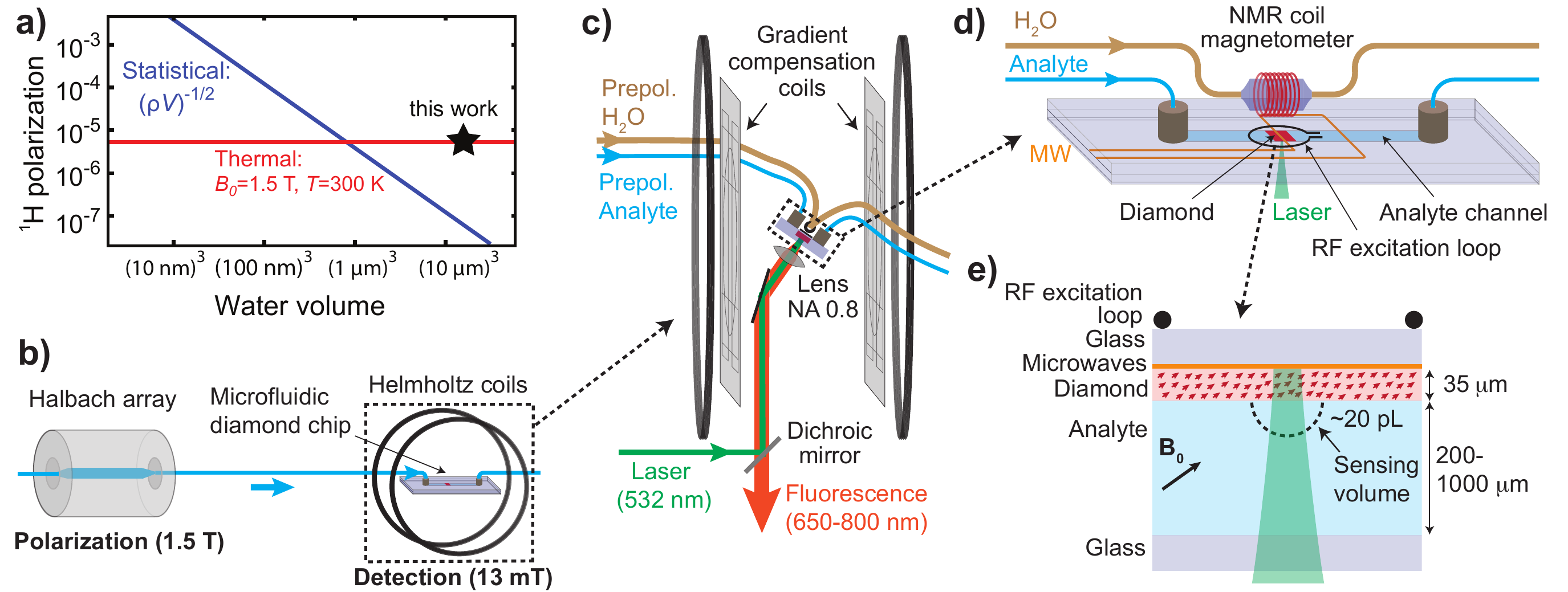}
    \caption{Microfluidic prepolarization NMR setup. a) Comparison of statistical and thermal polarization of protons in water as a function of detection volume. The room-temperature water proton density is $\rho=6.7{\times}10^{28}~{\rm m^{-3}}$. b) Prepolarization concept. Analyte is prepolarized by flowing it through a permanent magnet (1.5 T Halbach array). It is subsequently shuttled to a microfluidic chip housed in a stabilized, lower magnetic field ($B_0=13~{\rm mT}$, Helmholtz coils) where it is detected by NV NMR. c) Detection setup. Prepolarized analyte flows to a microfluidic chip where it is stopped, via fludic switches (not shown), and the NV NMR signal is detected using a custom-built epifluorescence microscope with $0.8$ numerical aperture (NA). A set of eight gradient compensation coils are used to eliminate first- and second-order magnetic field gradients along the field direction. The field is stabilized temporally using a coil-based NMR magnetometer in combination with low-inductance feedback coils wound around the main Helmholtz coils.  d) Microfluidic chip setup. The chip is constructed using milled glass and epoxy. Two fluidic lines pass to the detection region, one consisting of water (for NMR coil magnetometer) and the other with analyte (for NV NMR). A radio frequency (RF) excitation loop, placed in between the NMR coil magnetometer and the NV NMR sensor, excites nuclear spin coherence in both channels. The NMR coil magnetometer consists of a 3 mm diameter coil wound around a ${\sim}10~{\rm \upmu L}$ water volume. Copper microwave (MW) lines, printed on the interior of the glass chip, provide spin control over NV electron spins. e) NV NMR geometry. An NV-doped diamond membrane ($1{\times}1{\times}0.035~{\rm mm^3}$) is located on the surface of a microfluidic channel ($0.2\mbox{--}1{\times}2~{\rm mm^2}$ cross section) in contact with the analyte. Laser illumination (532 nm) bounces off the printed microwave line, and fluorescence (650-800 nm) is detected. The effective analyte detection volume is ${\sim}20~{\rm pL}$ (Sec.~\ref{sec:nmrsig}).}
\label{fig:exp_setup}
\end{figure*}

Quantum sensors based on nitrogen-vacancy (NV) centers in diamond have emerged as an alternative NMR detection modality due to their sub-micron spatial resolution and non-inductive based detection. Early implementations probed the nanoscale fluctuations of nuclear magnetization (statistical polarization) to enhance SNR \cite{staudacher2013nuclear,mamin2013nanoscale}. However, nanoscale diffusion of the analyte across the sensing volume broadened the spectral distribution to ${\sim}1~{\rm kHz}$, masking the informative spectral features arising from chemical shifts and $J$-couplings \cite{staudacher2015probing,kehayias2017solution}. The use of viscous solvents \cite{aslam2017nanoscale} improved the frequency resolution to ${\sim}100~{\rm Hz}$, enabling the resolution of large chemical shifts at $B_0=3~{\rm T}$. While further improvements in resolution are possible by increasing the detection volume ($V$), these come at a steep cost in SNR since statistical polarization scales $\propto V^{-1/2}$, Fig.~\ref{fig:exp_setup}(a). 

Alternatively, at sufficiently large $V$ and $B_0$, the net thermal (Boltzmann) polarization becomes the dominant contributor to nuclear polarization, see Fig.~\ref{fig:exp_setup}(a). Detection of thermal polarization was recently demonstrated \cite{glenn2018high} using an NV-based NMR spectrometer achieving a frequency resolution of ${\sim}10~{\rm Hz}$ operating at $B_0=88~{\rm mT}$. This resolution was sufficient to detect large spectral splittings due to proton chemical shifts and $J$-couplings, with a concentration sensitivity (defined throughout as the minimum detectable proton concentration for SNR=3) of ${\sim}370~{\rm M~s^{1/2}}$. 

In this manuscript, we report an order-of-magnitude improvement in spectral resolution, $0.65~\pm0.05~{\rm Hz}$, and realize a concentration sensitivity  of ${\sim}27~{\rm M~s^{1/2}}$. This is accomplished by spatially separating the polarization and detection phases of the experiment in a microfluidic setup \cite{ANO2001,LED2006}. Strong permanent magnets (1.5 T) are used to generate nuclear spin polarization. Detection is performed at 13 mT using Helmholtz coils, simplifying the task of stabilizing NMR linewidths to sub-Hz levels while enabling the use of high-sensitivity diamond quantum sensing protocols at low microwave frequencies \cite{glenn2018high}. These improvements allow us to perform two-dimensional (2D) correlation spectroscopy (COSY) of liquid analytes within an effective ${\sim}20$ picoliter detection volume. The use of diamond quantum sensors as in-line microfluidic NMR detectors is a significant step towards applications in mass-limited chemical analysis and single cell biology. In combination with advances in dynamic nuclear polarization using external polarizing agents \cite{bucher2018hyperpolarization} and, potentially, optical hyperpolarization using NV centers \cite{london2013detecting, king2015room,fernandez2018toward,ajoy2018orientation,pagliero2018multispin,broadway2018quantum}, this platform may eventually enable NMR spectroscopy of metabolites at physiological concentrations with single-cell spatial resolution.

\section{Experimental setup}
Figure \ref{fig:exp_setup}(b) illustrates the prepolarization concept. Fluid analytes are housed in a helium-pressurized container which enables variable flow rates up to $50~{\rm \upmu L/s}$. The analyte first flows through a 1.5 T Halbach array for a dwell time of ${\sim}6~{\rm s}$. This time is longer than the longitudinal spin relaxation time of the analytes studied here ($T_1\approx2~{\rm s}$), leading to an equilibrium polarization of ${\sim}5{\times}10^{-6}$. The analyte then flows to a detection region where it is detected by NV NMR. In order for the analyte to retain the thermal polarization generated in the prepolarization step, the transfer must be performed adiabatically (the rate of change in the magnetic field angle should be much smaller than the nuclear spin angular frequency) and on a shorter timescale than $T_1$ \cite{Tayler2017}. Both conditions are satisfied by ensuring that the analyte never passes through a magnetic field region smaller than ${\sim}0.3~{\rm mT}$ and by limiting the transfer time to ${\sim}0.5~{\rm s}<T_1$ (Sec.~\ref{sec:ad}). Microfluidic switches ensure that the analyte is transferred to the detection region and then stopped for NMR detection (see Sec.~\ref{sec:flowswitch}). 

Figure \ref{fig:exp_setup}(c) depicts the detection setup. Helmholtz coils produce a magnetic field $B_0=12.935~{\rm mT}$, corresponding to a proton resonance frequency $\gamma_p B_0=550.75~{\rm kHz}$, where $\gamma_p=42.577~{\rm MHz/T}$ is the proton gyromagnetic ratio. A set of gradient compensation coils, consisting of eight separate current-carrying wire configurations, enables cancellation of first- and second-order magnetic field gradients along the field direction (see Sec.~\ref{sec:gradcomp}). The magnetic field is temporally stabilized using a feedback loop incorporating a custom NMR coil magnetometer positioned just above the diamond detection volume, Fig.~\ref{fig:exp_setup}(d). Prepolarized water continuously flows through a $3~{\rm mm}$ diameter NMR detection coil. The water's proton nuclear precession is initialized by a $\pi/2$ pulse using the same radio frequency (RF) loop used for NV NMR. The inductively-detected coil signal is amplified, digitized, and fit for the proton NMR frequency. The instantaneous magnetic field is inferred, and temporal deviations are actively compensated by altering the current in a pair of low-inductance compensation coils. With this system, we realize a temporal field stability of ${\sim}1~{\rm ppm}$ (${\sim}0.6~{\rm Hz}$ at the proton NMR frequency), limited by the accuracy of the NMR coil magnetometer (Sec.~\ref{sec:temporal}). 

The microfluidic chip housing the diamond sensor is depicted in Fig.~\ref{fig:exp_setup}(d). The components of the chip include a copper loop (printed on a glass slide) used to deliver microwaves, an RF excitation loop placed between the diamond and the feedback NMR coil, a microfluidic channel enclosing the diamond sensor and contacting analyte, and microfluidic ports to mate the external analyte tubing with the chip. An enlarged picture of the chip surrounding the diamond sensor is shown in Fig.~\ref{fig:exp_setup}(e). A $20~{\rm \upmu m}$ diameter laser beam excites NV centers throughout a $35~{\rm \upmu m}$ thick diamond membrane. Magnetostatic modeling indicates that $50\%$ of the NMR signal comes from a $20~{\rm pL}$ hemispherical region of analyte above the optical axis (Sec.~\ref{sec:nmrsig}). By convention \cite{glenn2018high}, we defined this region as the effective detection volume. Several $35~{\rm \upmu m}$ thick diamond membranes were used with $1{\times}1~{\rm mm^2}$, [100] polished faces. The membranes were formed from diamond chips grown by either high pressure high temperature synthesis or chemical vapor deposition and hosted an initial nitrogen density of $20\mbox{--}50~{\rm ppm}$. The chips were irradiated with 2 MeV electrons at a dose of ${\sim}10^{18}~{\rm cm^{-2}}$ and subsequently annealed at $800\mbox{--}1100^{\circ}~{\rm C}$ using the recipe described in \cite{kehayias2017solution}. NV centers in the processed membranes exhibit a coherence time of $10\mbox{--}20~{\rm \upmu s}$ under an XY8-1 pulse sequence. 

NV NMR detection was performed using a custom-built epifluorescence microscope, Fig.~\ref{fig:exp_setup}(c). Linearly-polarized pulses of laser light (0.3 W, 532 nm) polarize and detect the spin projection states of NV centers, via their spin-dependent fluorescence. The fluorescence is spectrally filtered (650-800 nm) and imaged onto a photodetector, producing ${\sim}10~{\rm \upmu A}$ of peak photocurrent. The diamond membranes are oriented so that one of the four possible NV axes is aligned with the magnetic field. The ODMR transitions of these aligned NV centers is $D\pm\gamma_{NV}B_0$, where $D=2.87~{\rm GHz}$ is the NV zero field splitting and $\gamma_{NV}=28.0~{\rm GHz/T}$ is the NV gyromagnetic ratio. NV center spin states are manipulated using microwaves resonant with the lower-frequency transition, $2.51~{\rm GHz}$. Throughout, we set the microwave power to produce a $\pi$ pulse length of 44 ns, and the fluorescence-detected contrast of Rabi oscillations was typically $8\%$ peak-to-peak.

\begin{figure}[ht]
\centering
\includegraphics[width=\columnwidth]{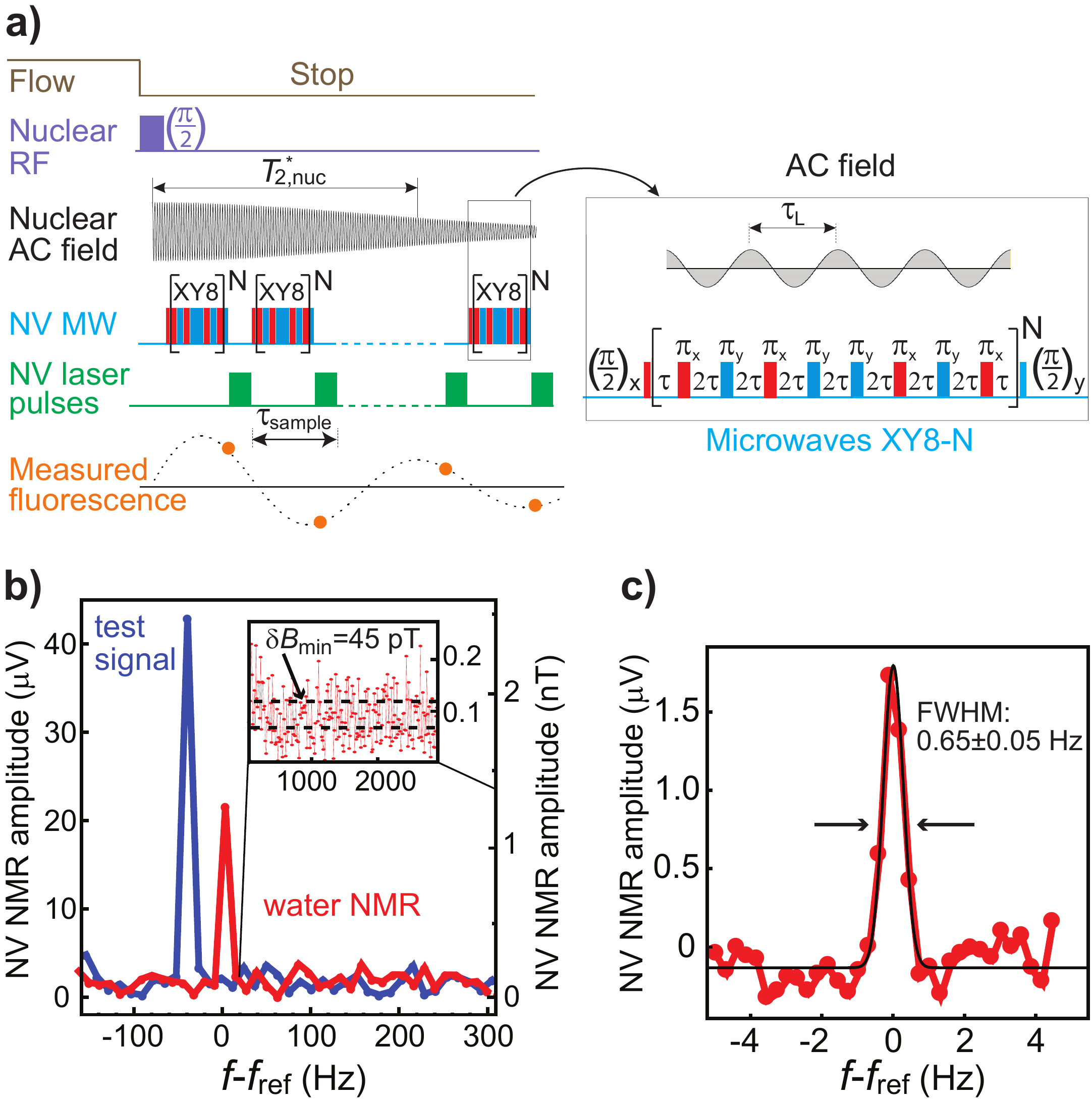}
\caption{Characterization of prepolarized NV NMR. a) The synchronized readout pulse sequence. It consists of a train of XY8-N pulses that perform successive phase measurements of the AC magnetic field produced by precessing nuclei. The measured fluorescence reflects an aliased version of the nuclear AC field. The entire sequence is repeated every $2.5\mbox{--}4.25~{\rm s}$ ($1.25~{\rm s}$ for flow, the remainder for detection). b) NV NMR spectra (absolute value of Fourier transform) of water (red) and an applied $2.5~{\rm nT}$ amplitude test field (blue) for an effective acquisition time of $5.2~{\rm s}$ (excluding dead time). Inset: the standard deviation of the noise floor reveals $\delta B_{\rm min}=45~{\rm pT}$. From these data we infer a minimum detectable concentration of $27~{\rm M~s^{1/2}}$ (SNR=3). Incorporating all experimental dead time, the concentration sensitivity is ${\sim}45~{\rm M~s^{1/2}}$, Sec.~\ref{sec:concsen}. c) A high-resolution NV NMR spectrum of water (imaginary part of Fourier transform) reveals a FWHM linewidth of $0.65\pm0.05~{\rm Hz}$. The data were obtained by averaging 60 traces, each $3~{\rm s}$ long. 
}
\label{fig:pulse_seq_and_SNR}
\end{figure}

\begin{figure*}[ht]
\centering
\includegraphics[width=0.9\textwidth]{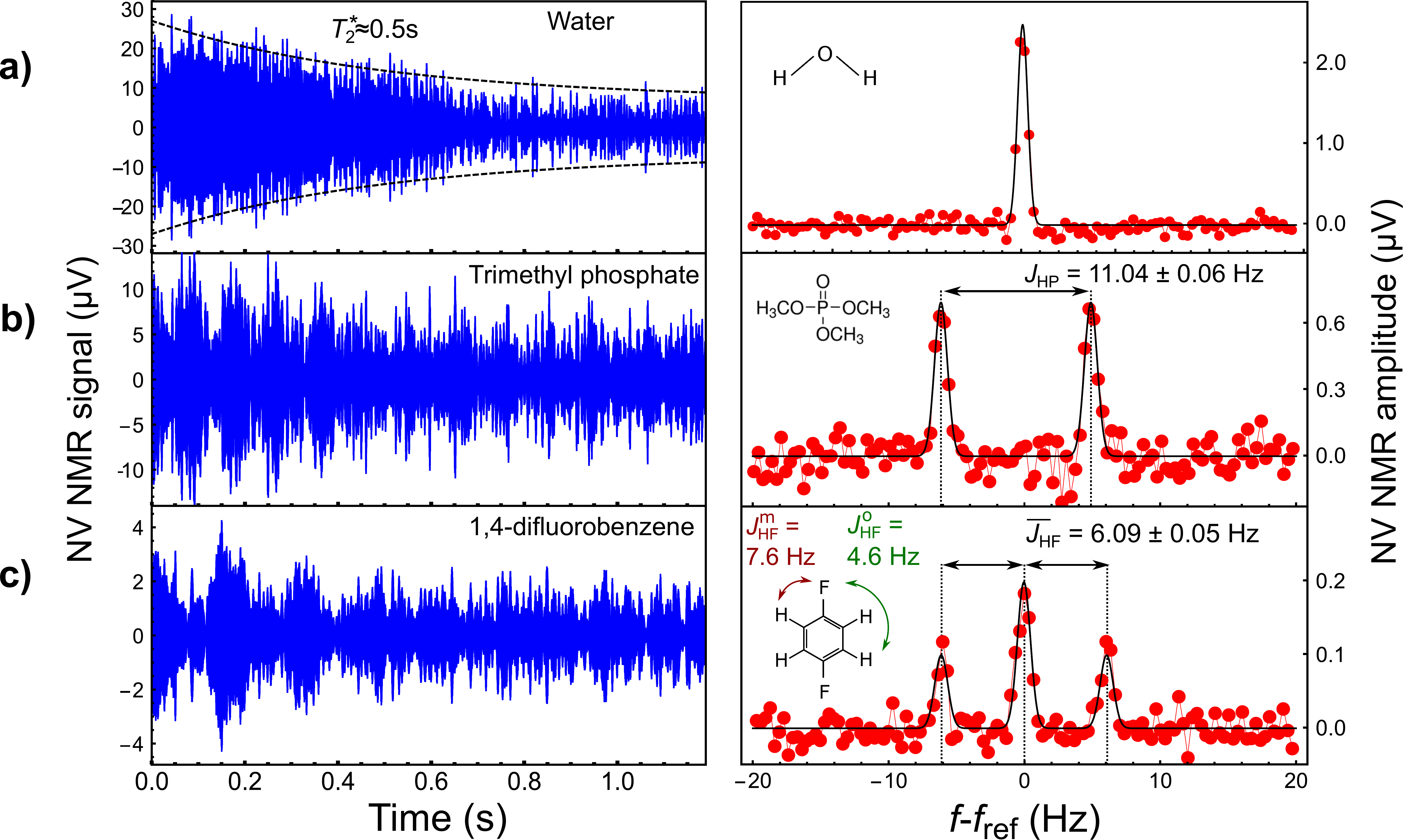}
\caption{One-dimensional NMR. Time-domain (left) and frequency-domain (right) NV NMR signals for a) water, b) trimethyl phosphate (TMP) and c) 1,4-difluorobenzene (DFB). Signals were averaged over ${\sim}10^3$ traces for a total of ${\sim}1~$hour acquisition. A ${\sim}1~{\rm kHz}$ bandwidth bandpass filter is applied to the time-domain data for better visualization. The frequency-domain spectra show the imaginary component of the Fourier transform. Each spectrum is fit with Gaussian functions (black lines). For TMP, we constrain the widths of both lines to be equal with a 1:1 amplitude ratio and find $J_{\rm HP}=11.04\pm0.06~{\rm Hz}$. For DFB, we constrain the widths of all three lines to be equal with a 1:2:1 amplitude ratio and find $\overline{J_{\rm HF}}=6.09\pm0.05~{\rm Hz}$.
} 
\label{fig:1Dspectra}
\end{figure*}

\begin{figure*}[ht]
    \centering
    \includegraphics[width=0.85\textwidth]{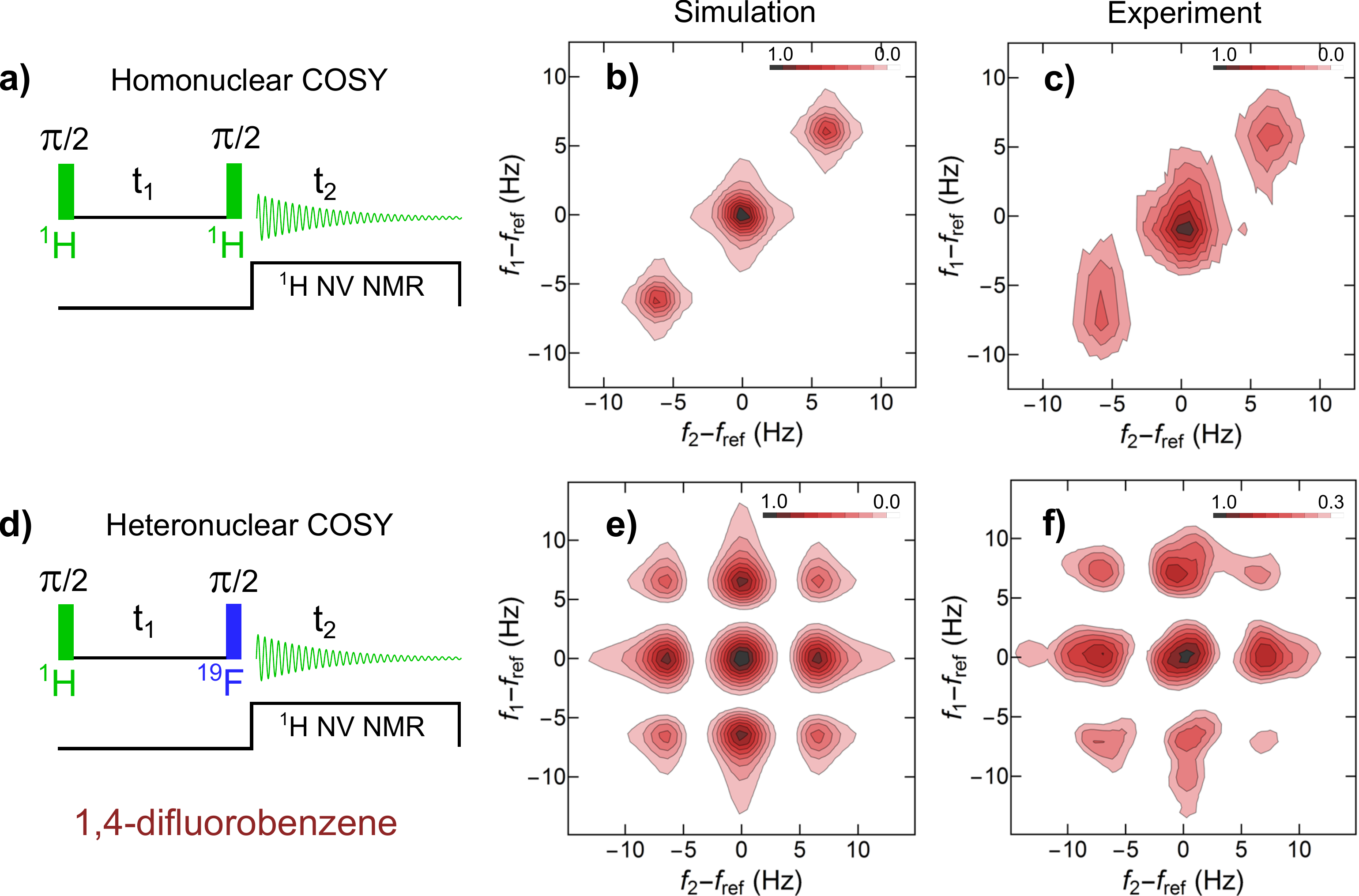}
    \caption{Two-dimensional COSY NMR of 1,4-difluorobenzene (DFB). a) Homonuclear COSY pulse sequence, b) simulated spectrum, and c) experimental NV NMR spectrum of DFB. d) A modified heteronuclear COSY sequence reveals off-diagonal peaks in both e) simulation and f) experiment. Color scales correspond to the normalized absolute value of the 2D Fourier transform. Vertical axes ($f_1-f_{\rm ref}$) correspond to the frequencies of the $t_1$ dimension, and horizontal axes ($f_2-f_{\rm ref}$) correspond to the frequencies of the $t_2$ dimension. In c), 14 values of $t_1$ in 0.021 s increments up to 0.294 s were used. Total acquisition time was $22~$hours.
    In f), 16 values of $t_1$ in 0.021 s increments up to 0.336 s are used. Total acquisition time was $25~$hours. In both cases the $t_2$ acquisition spanned from 0 to 1.25 s. All simulations were performed using the SPINACH package \cite{hogben2011spinach}. Simulation and experimental data use the same windowing functions (see Sec.~\ref{sec:spinach}). 
    }
    \label{fig:2D_dfb}
\end{figure*}

The pulse sequence used to detect NV NMR is depicted in Fig.~\ref{fig:pulse_seq_and_SNR}(a). It shares common traits with the synchronized readout scheme used in Refs.~\cite{glenn2018high,bucher2018hyperpolarization}. A $\pi/2$ RF pulse (${\sim}1~{\rm ms}$ long), resonant with the proton spin transition, initializes nuclear spin precession, producing an exponentially-decaying oscillating (AC) magnetic field with a nominal frequency $f_{\rm ref}=1/\tau_L=550.75~{\rm kHz}$. Subsequently, a series of XY8-5 microwave pulse sequences are applied to the NV centers to detect the nuclear AC field. Only the component of the nuclear field along the NV axis is detected \cite{glenn2018high}. Each XY8-5 sequence contains forty $\pi$ pulses separated by $\tau_L/2$. After each XY8-5 sequence, a $3.4~{\rm \upmu s}$ laser pulse is applied to the NV centers for optical readout and repolarization. The first $0.5~{\rm \upmu s}$ of the readout fluorescence is used to measure the NV spin projection and the final $1~{\rm \upmu s}$ is used for normalization to eliminate low-frequency intensity noise. 

Each NV readout nominally measures the initial phase of the nuclear AC field. A time series of the NV readouts yields an aliased version of the nuclear AC field with frequency $f_{\rm alias}=f_{\rm ref}-f_{\rm sample}{\times}{\rm Round}(f_{\rm ref}/f_{\rm sample})$, where
$f_{\rm sample}=1/\tau_{\rm sample}\approx24~{\rm kHz}$ is the sampling frequency of NV readouts. Unlike the sequence used in Ref.~\cite{glenn2018high}, the duration of each XY8-5 sequence is held constant at the point of maximal sensitivity, while $f_{\rm alias}$ is varied, up to the maximum frequency $f_{\rm sample}/2\approx12~{\rm kHz}$, by adjusting a small dead time between readouts. 

\section{Sensitivity and spectral resolution}
The sensitivity and spectral resolution limits of our apparatus were determined from measurements on de-ionized water. Figure~\ref{fig:pulse_seq_and_SNR}(b) shows results of the sensitivity measurements. An AC magnetic field, with a calibrated amplitude of 2.5 nT (see Sec.~\ref{sec:magcal}), was detuned slightly from $f_{\rm ref}$ and detected using the NV NMR pulse sequence. The NV NMR signal from prepolarized water was then recorded under identical conditions. The Fourier transform of the water signal reveals an amplitude of 1.21 nT. Magnetostatic modeling predicts that a proton polarization of $3.9{\times}10^{-6}$ would produce this signal strength (Sec.~\ref{sec:nmrsig}). This signifies that $77\%$ of the maximum thermal polarization generated in the Halbach array ($5.1{\times}10^{-6}$ for 1.5 T at 300 K) is retained. The standard deviation of points near the resonance peak [inset of Fig.~\ref{fig:pulse_seq_and_SNR}(b)] reveals a magnetic noise of $0.10~{\rm nT~s^{1/2}}$. This corresponds to a concentration sensitivity of $27~{\rm M~s^{1/2}}$ for SNR=3. Between experiments, the concentration sensitivity varied by ${\sim}50\%$ depending on the fluorescence level, contrast, and NV coherence time of the diamond illumination region.

To optimize the spectral resolution, the gradient compensation coils were adjusted until no perceptible decrease in NV NMR linewidth was observed. Figure \ref{fig:pulse_seq_and_SNR}(c) shows an NV-detected water NMR spectrum with one of the narrowest linewidths obtained. A Gaussian fit reveals a full-width-at-half-maximum (FWHM) of $0.65 \pm 0.05 $ Hz. While a substantial improvement over previous studies, this is broader than the expected natural linewidth of water under our experimental conditions, ${\sim}0.1~{\rm Hz}$~\cite{HAR2010}. We attribute the discrepancy to residual temporal instability in $B_0$ (Sec.~\ref{sec:temporal}). 

\section{1D NMR}
To showcase the capabilities of our NV NMR spectrometer, we obtained proton NMR spectra of different fluid analytes. Figure \ref{fig:1Dspectra}(a) shows the time and frequency domain signals of water. The SNR is sufficient to resolve the decay in the envelope of the proton magnetization, from which we infer a spin dephasing time $T_2^{*}\approx0.5~{\rm s}$, consistent with the sub-Hz linewidths observed in the frequency domain. 

Figure \ref{fig:1Dspectra}(b) shows the NV NMR spectrum for trimethyl phosphate (TMP). The characteristic beats in the time domain and spectral splitting in the frequency domain are signatures of $J$-coupling. Such splittings arise due to terms in the nuclear spin Hamiltonian of the form $J_{12} \vec{I_1}\cdot \vec{I_2}$, where $\vec I_{1}$ and $\vec I_{2}$ are the spin angular momenta of different nuclei. At $B_0=13~{\rm mT}$, couplings between spins of different isotopes (``heteronuclear'' $J$-coupling) lead to well-defined splittings in the NMR spectra, whereas homonuclear $J~$splittings are not resolved \cite{appelt2007phenomena}. The $11.04\pm0.06~{\rm Hz}$ splitting in the TMP spectrum corresponds to the known heteronuclear $J$-coupling between the $^{31}$P nuclear spin and each of the equivalent $^1$H spins \cite{liao2010study}.

Figure~\ref{fig:1Dspectra}(c) shows the NV NMR spectrum for 1,4-difluorobenzene (DFB). In DFB (inset), each proton is coupled to the nearest $^{19}\rm F$ atom with $J_{\rm HF}^{\rm m}=7.6~{ \rm Hz}$ and the further $^{19}\rm F$ atom with $J_{\rm HF}^{\rm o}=4.6~{\rm Hz}$ \cite{paterson1964nmr}. The spectrum exhibits an average of the two splittings, $\overline{J_{\rm HF}}=6.09\pm0.05~{\rm Hz}$, with a 1:2:1 amplitude ratio, consistent with previous reports \cite{robinson2006two, paterson1964nmr}.

\section{2D NMR}
Having established the ability to detect NMR spectra with sub-Hz resolution and high SNR, we next used our platform to perform 2D COSY NMR spectroscopy. Multidimensional NMR spectroscopy enables the determination of nuclear interactions within complex structures, even in cases where the corresponding 1D spectra are complicated or have ambiguous interpretation. It is widely used in applications ranging from metabolomics to protein structure identification \cite{kovermann2016protein,markley2017future}.

We performed two different variations of the 2D COSY experiment that probe the nuclear interactions within DFB. In the first case, homonuclear COSY \cite{aue1976two}, shown in Fig.~\ref{fig:2D_dfb}(a), two $\pi/2$ pulses on the proton spins are separated by a variable evolution period, $t_1$. Following the second pulse, the precessing proton magnetization is continuously recorded as a function of time, $t_2$. The sequence is then iterated by incrementing $t_1$ to build up a 2D array. 

Figure \ref{fig:2D_dfb}(b,c) shows the 2D Fourier transform of the resulting array for DFB alongside a simulated spectrum obtained by density matrix modeling (Sec.~\ref{sec:spinach}) using the SPINACH software package \cite{hogben2011spinach}. The presence of three diagonal peaks separated by 6.1 Hz indicates that the proton magnetization is modulated at the heteronuclear $J$-coupling frequency during the $t_1$ evolution interval. However, the absence of cross peaks indicates a lack of magnetization transfer between the spin states. This is expected since there is no difference in the chemical shift between the protons~\cite{paterson1964nmr}. A homonuclear COSY spectrum of TMP is presented in Sec.~\ref{sec:tmp2d}.
 
In the second 2D NMR experiment on DFB, we used a modified heteronuclear COSY sequence where the second $\pi/2$ pulse is resonant with $^{19}\rm F$ nuclei ($518.08~{\rm kHz}$), Fig.~\ref{fig:2D_dfb}(d). As before, the pulses are separated by a variable evolution time, $t_1$, and we tune our NV NMR sequence to selectively detect the proton precession as a function of $t_2$. The simulated and experimental 2D Fourier transforms are shown in Fig.~\ref{fig:2D_dfb}(e,f). The presence of cross peaks separated by ${\sim}6~{\rm Hz}$ indicates that the $^{19}\rm F$ pulse mediates transfer of magnetization amongst the $J$-split proton spin states. The results are consistent with previous findings on DFB at Earth's magnetic field \cite{robinson2006two}. In Sec.~\ref{sec:2dsim} we provide an analytical calculation of a two spin model which effectively describes these dynamics.

\section{Outlook and conclusion}
\label{sec:outlook}
The demonstration of sub-Hz resolution and multidimensional NMR paves the way for diamond quantum sensors to be used in applications such as in-line hyphenated analysis \cite{BRK2011}, single-cell metabolomics \cite{ZEN2013}, and mass-limited pharmacodynamics \cite{markley2017future}. The high spatial resolution, epifluorescence imaging format of our sensor lends itself to parallelization, which could enable high throughput chemical analysis or NMR imaging of cell cultures with single cell resolution.

A remaining challenge is that the present sensor would require substantial averaging times for detection of metabolites at physiological concentrations ($\rm \upmu M\mbox{--}mM$). In the short term, up to an order-of-magnitude improvement in NV NMR sensitivity may be realized by detecting at higher magnetic field (which would enable the use of longer, more sensitive XY8-N sequences) \cite{DEL2011}, improving the photon collection efficiency \cite{SIY2010,LES2012}, and increasing the NV emission intensity and contrast through optimized diamond doping \cite{ACO2009,KLE2016,ISH2017}. Another order of magnitude improvement in concentration sensitivity is possible by using a superconducting magnet for prepolarization \cite{VER2008}. The use of external polarizing agents may improve the sensitivity by up to two orders of magnitude \cite{bucher2018hyperpolarization}, provided that such additives are compatible with the target assay. In the longer term, the largest gains in sensitivity may come from the use of optical hyperpolarization methods to transfer the near-unity NV electron spin polarization to the analyte non-invasively \cite{london2013detecting, king2015room,fernandez2018toward,ajoy2018orientation,pagliero2018multispin,broadway2018quantum}.

In summary, we demonstrated that diamond quantum sensors can be used in microfluidic NMR applications. We showed that separating polarization and detection steps enabled an order-of-magnitude improvement in spectral resolution (0.65 Hz) over previous diamond NMR studies, with a concentration sensitivity of ${\sim}27~{\rm M s^{1/2}}$. We used the platform to perform two-dimensional NMR on fluid analytes and observed the transfer of magnetization mediated by heteronuclear $J$-coupling. 

\begin{acknowledgments}
The authors acknowledge valuable conversations with C. Avalos, V. Bajaj, A. Pines, D. Budker, J. Blanchard, R. Walsworth, D. Bucher, and A. Ramamoorthy. This study was funded by NIH (NIGMS) award 1R41GM130239-01 and a Beckman Young Investigator award. J. Smits acknowledges support from the Latvian Government (A5-AZ27, Y9-B013).

\textbf{Competing interests.} A. Jarmola, and V. M. Acosta are co-inventors on a related patent: US 2018/0203080 A1. A. Jarmola is a co-founder of startup ODMR Technologies and has financial interests in the company. The remaining authors declare no competing interests.

\textbf{Author contributions.} A. Jarmola and V. M. Acosta conceived the initial idea for the experiment in consultation with P. Kehayias and J. Smits. J. Smits, J. Damron, A. Jarmola, and V. M. Acosta designed the experiment with input from all authors. A. F. McDowell provided expertise and hardware for field stabilization. J. Smits and P. Kehayias wrote and implemented control and automation software. J. Smits, J. Damron, A. Jarmola, and V. M. Acosta constructed the apparatus, performed the experiments, and analyzed the data, with assistance from I. Fescenko, A. Laraoui, N. Mosavian, and N. Ristoff in data acquisition and microfluidic sensor fabrication. All authors discussed results and contributed to the writing of the manuscript.
\end{acknowledgments}

\bibliographystyle{apsrev4-1}
\bibliography{ref} 

\widetext
\clearpage

\begin{center}
\textbf{\large Supplemental Information: Two dimensional nuclear magnetic resonance spectroscopy with a microfluidic diamond quantum sensor}
\end{center}
\setcounter{equation}{0}
\setcounter{section}{0}
\setcounter{figure}{0}
\setcounter{table}{0}
\setcounter{page}{1}
\setcounter{equation}{0}
\setcounter{figure}{0}
\setcounter{table}{0}
\setcounter{page}{1}
\makeatletter
\renewcommand{\thetable}{S\arabic{table}}
\renewcommand{\theequation}{S\arabic{equation}}
\renewcommand{\thefigure}{S\arabic{figure}}
\renewcommand{\thesection}{S\Roman{section}}
\renewcommand{\bibnumfmt}[1]{[S#1]}
\renewcommand{\citenumfont}[1]{S#1}

\section{NV NMR detection apparatus}
A 532 nm green laser (Lighthouse Photonics Sprout G-10W) beam was used to excite NV centers. An aspheric lens (Thorlabs ACL12708U, NA$\approx0.8$) focused the $0.3~{\rm W}$ laser beam to illuminate a $20~{\rm \upmu m}$ diameter patch of diamond. Red fluorescence was separated from the excitation light by a dichroic mirror and detected by an amplified photodetector (Thorlabs PDB450A) with $4~{\rm MHz}$ bandwidth. A HighFinesse Gmbh UCS 10/40 ultra-low noise current source was used to drive the Helmholtz coils with 9.4 A. 

The experiment was controlled by a TTL pulse card (PBESR-PRO-500 by SpinCore). Laser pulses were generated by passing the continuous-wave laser beam through an acousto-optic modulator (AOM, CrystaLaser). The modulator was driven by a 100 MHz RF source (Trinity Power TPI-1001-B). The source's amplitude was modulated on a $\lesssim10~{\rm ns}$ timescale using a switch (Mini-Circuits ZASWA-2-50DR). It was subsequently amplified to $1~{\rm W}$ and delivered to the AOM. 

Microwave pulses were generated using an I/Q modulated microwave generator (SRS SG384). The microwave amplitude and phase were controlled on a $\lesssim10~{\rm ns}$ timescale using a series of TTL controlled switches. The NV $\pi$ pulse length was set to 44 ns and $\pi/2$ pulse length was 22 ns. The microwaves subsequently passed through an amplifier (Mini-Circuits ZHL-16W-43+) and circulator and were connected to the NV NMR chip.

RF pulses used for NMR $\pi/2$ pulses were generated by an arbitrary waveform generator (AWG, Teledyne LeCroy WaveStation 2012). The pulse timing was controlled by a trigger pulse from the TTL pulse card. A pulse signal from the AWG was also used for synchronizing the timing of the microfluidic flow switches. NMR $\pi/2$ pulses were typically ${\sim}1~{\rm ms}$ long and applied on nuclear spin resonance (550750 Hz for $^1\rm H$ and 518082 Hz for $^{19}\rm F$). 

Two data aquisition (DAQ) cards (NI USB-6361) were used. One DAQ was used for NV NMR data acquisition, and the other was used for the NMR coil magnetometer and temporal feedback. The clocks of the two DAQs and TTL pulse card were synchronized using low-drift oven-controlled crystal oscillators. For NV NMR, the photodetector signal was digitized and sychronized to the overall pulse sequence by using the TTL pulse card to generate the sample clock for the NV NMR DAQ.

\section{Magnetic field gradient compensation}
\label{sec:gradcomp}

The magnetic field was stabilized spatially using a set of eight gradient compensation coils (built by NuevoMR, LLC.), driven by two 4-channel power supplies (Instek GPD-4303S) producing currents in the $0\mbox{--}0.5~{\rm A}$ range. The compensation coils were constructed according to the design described in Ref. \cite{Anderson1961}. Three channels compensated the linear field gradients, and the other 5 channels compensated second-order gradients. The design was modified slightly by replacing the $X^2-Z^2$ and $Y^2-Z^2$ coils with $Z^2$ $ (=3Z^2-X^2-Y^2)$ and $X^2-Y^2$. Here the coils are labeled according to the Cartesian representation of the spherical harmonic terms in the field expansion, Ref. \cite{Anderson1961}.  The spacing between the shim planes was $80~{\rm mm}$. Five turns of 24 AWG enameled magnet wire were laid out into double sided adhesive tape attached to 0.005" thick FR4 fiberglass boards, following a template visible through the FR4. Multiple coils were laid out and then sealed under epoxy, with a separate FR4 layer on top. The coils produce correction fields of $1.2\mbox{--}1.6~{\rm \upmu T/mm/A}$ for the linear terms, and $0.05-0.1~{\rm \upmu T/mm^2/A}$ for the second-order terms.

\section{Gradients due to magnetic susceptibility mismatch of sensor components}
 \begin{figure}[ht]
    \centering
    \includegraphics[width=0.99\textwidth]{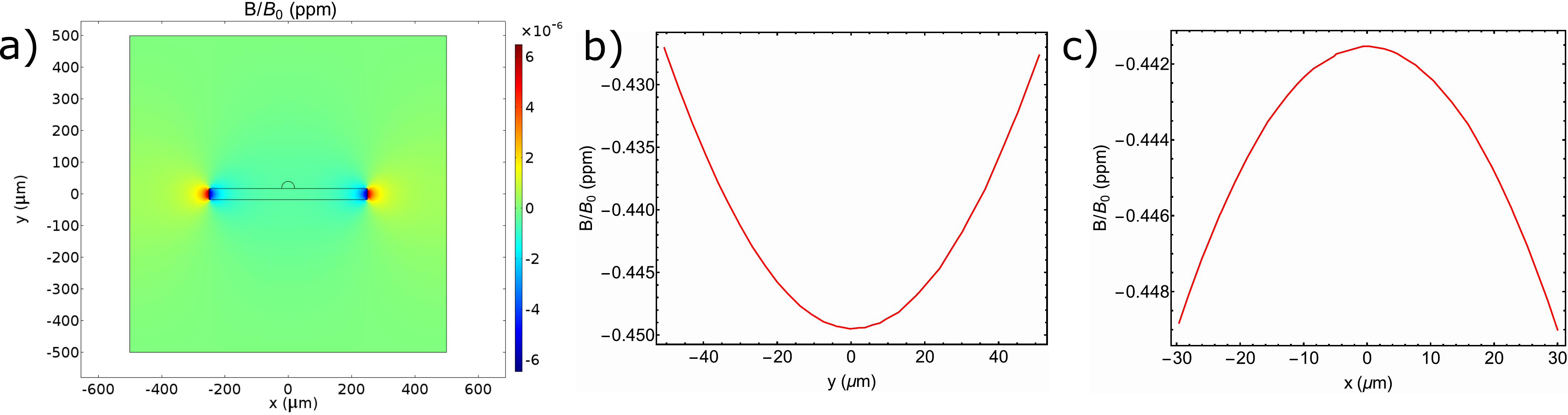}
    \caption{a) 2D map of the relative magnetic field around a diamond immersed in water. b) Line cut of the relative magnetic field at $x=0$. c) Line cut of the relative magnetic field for the value of $y$ located $10~{\rm \upmu m}$ above the diamond.
    }
    \label{fig:susceptibility_plots}
\end{figure}

To estimate the contribution of susceptibility mismatches between the sensor and analyte, we performed magnetostatic simulations on a simplified two-dimensional model of the sensor, Fig.~\ref{fig:susceptibility_plots}(a). A $35{\times}150~{\rm \upmu m^2}$ diamond membrane, with volume susceptibility $\chi_{V} \approx 2.2{\times}10^{-5}$, was placed in the center of a $1{\times}1~{\rm mm^2}$ volume of water (volume susceptibility $\chi_{V} \approx 9.6 \times 10^{-6}$). Constant inward flux density boundary conditions were imposed on the boundaries at $y=\pm 500 \upmu$m and zero flux conditions were imposed on the remaining two boundaries. The relative change in magnetic field of the entire modeled region is shown in Fig.~\ref{fig:susceptibility_plots}(a). Line cuts of the relative field through the analyte detection region are shown in Fig.~\ref{fig:susceptibility_plots}(b,c).

While the fringe field near the edges of the diamond produce a relative change in field of a few parts per million, near the center of the diamond, the field variation is only at the ppb level. Since our analyte detection region is near the center of a much larger diamond chip, we therefore do not expect susceptibility mismatch to play a significant role in the NMR line broadening. 

\section{NMR coil magnetometer feedback system}
\label{sec:temporal}

For temporal feedback, the NMR coil magnetometer was tuned to proton NMR resonance using an LC circuit. The coil resonator had a quality factor of $\sim20$ and an impedance of ${\sim}10~{\rm kHz}$. The output passed through a low-noise current amplifier (SRS 560) and was digitized by the  feedback DAQ. 

After each experiment with NV detection, a proton $\pi/2$ pulse was applied, and the NMR coil magnetometer signal was digitized and converted to a power spectrum. The water NMR central frequency was calculated by fitting the power spectrum. The field was actively stabilized temporally by delivering $0\mbox{--}0.5~{\rm A}$ of current to a pair of coils wound around the main Helmholtz coils. This current was delivered by a Thorlabs LDC220C current source, with its instantaneous current output controlled via an external analog modulation input. After each NMR coil magnetometer acquisition, the current in the feedback coils was adjusted to minimize detuning from the target central frequency.

 \begin{figure}[hb]
    \centering
    \includegraphics[width=0.4\textwidth]{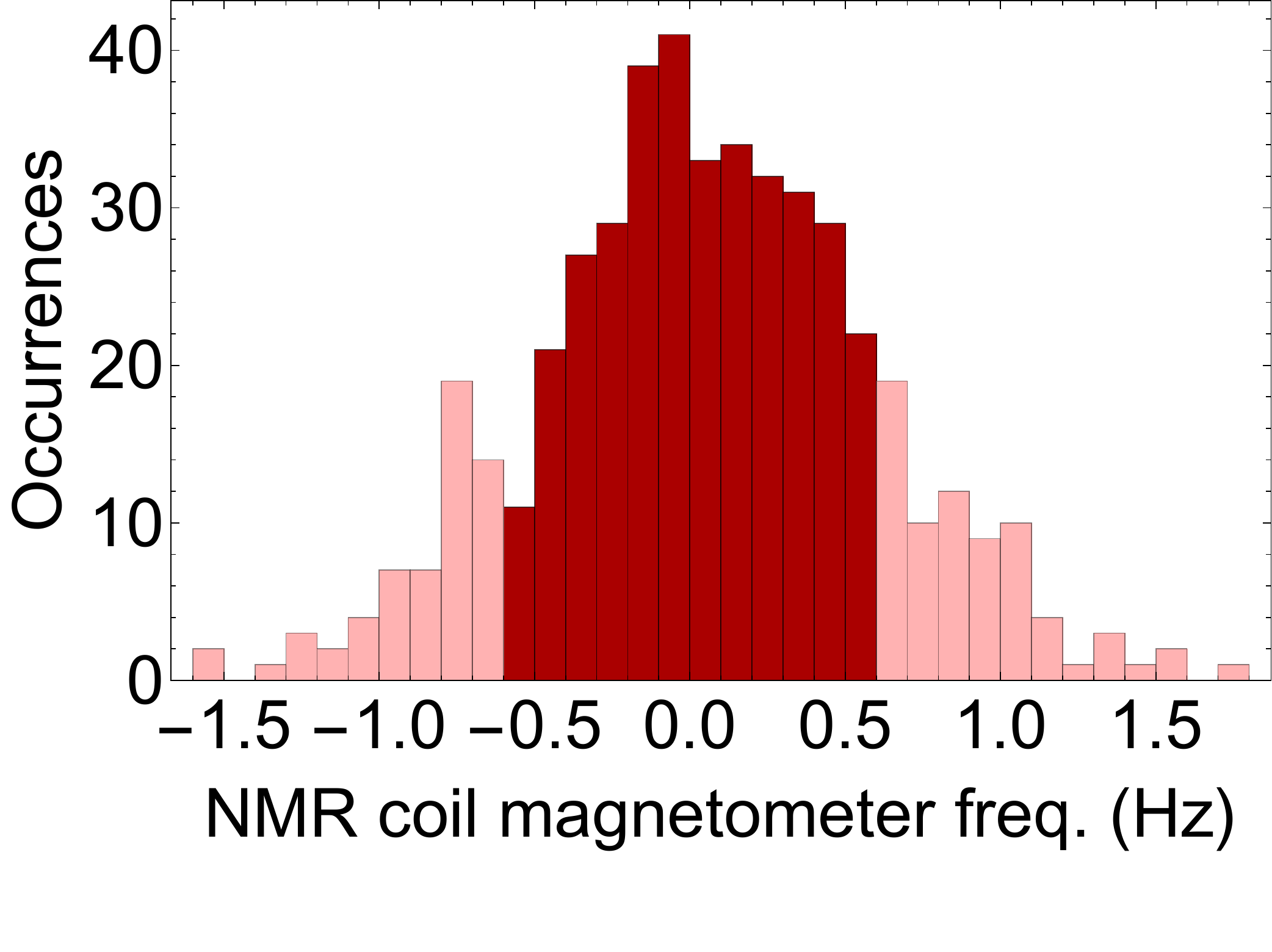}
    \caption{Histogram of fitted central frequencies obtained from the NMR coil magnetometer for a typical measurement. Whenever the observed deviation is larger than a threshold value ($\pm0.6~{\rm Hz}$ in this case) the corresponding NV NMR measurement is discarded.
    }
    \label{fig:temporal_drift}
\end{figure}

 During 1D and 2D NV NMR measurements, the deviation of the feedback value from the desired value was monitored and whenever it was larger than a threshold value the corresponding NV measurement would be discarded. We typically set the threshold so that $\lesssim25\%$ of total data was discarded. A histogram of NMR magnetometer frequency values obtained over the course of a few hours is shown in Fig.~\ref{fig:temporal_drift}. The spread of deviations in field values is consistent with the $0.5\mbox{--}1~{\rm Hz}$ linewidths observed in NV NMR experiments. The spread in values is due to a combination of (i) uncertainty in the fitted NMR magnetometer peak positions (${\sim}\pm0.3~{\rm Hz}$) and (ii) real changes in field that occur on a faster timescale than the feedback bandwidth (${\sim}0.3~{\rm Hz}$).

\section{Microfluidic Chip Fabrication}
The top surface of the microfluidic chips used in this study consisted of a 1-mm-thick microscope slide. Copper loops were fabricated on the slide and connected via a non-magnetic SMA solder jack to deliver microwaves. Two holes (an inlet and outlet), $0.5~{\rm mm}$ in diameter, were then drilled into the slide, using a diamond-tipped drill bit. These inlet/outlet holes were used to deliver fluid analytes to and from the chip. The diamond was then glued on top of the copper loops and oriented such that one of the NV axes would be aligned along the Helmholtz magnetic field once positioned in the setup. 

The microfluidic channel was defined by a spacer layer constructed in one of two ways. In the first method, a second 1-mm-thick microscope slide served as the walls of the microfluidic channel. The slide was cut to produce a ${\sim}35~{\rm mm}$ long channel spanning the inlet hole, diamond, and outlet hole. A slight taper was introduced at each end, with the widest part of the channel (${\sim}2~{\rm mm}$) in the center where the diamond was positioned. In the second method, a channel with similar length and width was cut from several layers of double-sided tapes (UltraTape 1510). The latter method enabled construction of thinner channels ($0.2\mbox{--}1~{\rm mm}$ thick).

The spacer layer was then either glued (in the case of the glass slide spacer) or adhered onto the copper coated slide with the affixed diamond. A $0.1\mbox{--}0.2~{\rm mm}$ thick coverslip was glued or adhered to the top of the spacer layer to seal the channel. The choice of two channel fabrication strategies arose from practical reasons associated with the analytes. Both TMP and DFB diffuse through and dissolve the plastic double-sided tapes. This can be partially mitigated by coating the channel with epoxy, a strategy we employed for experiments on TMP. However the chemical resistance was insufficient for long measurements on DFB, so we used the glass spacer design for experiments with DFB. A two-component Gorilla Epoxy was used for all gluing steps. 

Holes were drilled into two rubber stoppers to accommodate the 1/16 inch outer diameter PEEK tubing used for analytes. The stoppers were then attached with double sided tape (cutting a small hole in the tape for the tube opening) to the copper coated slide ensuring the tubing was aligned with the inlet/outlet holes. To reinforce the position and ensure a good seal, pressure was then applied to the stoppers using a piece of PCB board screwed into the Aluminum housing built for the entire chip. 

On the inlet stopper, a second hole orthogonal to (and slightly offset from) the analyte PEEK tubing hole was drilled. The prepolarized water sample used for the feedback NMR coil magnetometer was fixed to this hole in such a way that the magnetometer detection volume was displaced ${\sim}3~{\rm mm}$ from the diamond NV NMR detection region. Finally, a single-turn RF excitation loop was glued to the inlet stopper such that it lied between the diamond and NMR coil magnetometer. This loop was used to simultaneously excite the analyte in the microfluidic chip and the water running through the NMR coil magnetometer. 

\section{Sample Preparation}
Both trimethyl phosphate ($99\%$) and 1,4-difluorobenze ($99\%$) were purchased from Sigma Aldrich. All water samples were deionized before measuring. All analytes were degassed in a sonicator for a minimum of 0.5 hours before measuring. Water was typically degassed for roughly 2 hours at an elevated temperatures ($40\mbox{--}50^{\circ}~{\rm C}$), while trimethyl phosphate and difluorobenzene were kept at room temperature.

\section{Microdluidic Flow and switch timing}
\label{sec:flowswitch}
 After sonication, the analyte was placed in a 600 mL glass container with a screwtop lid that was pressurized under helium to drive flow. The container was pressurized anywhere from 60 to 100 PSI to achieve the desired flow rates, which varied from analyte to analyte. PEEK tubing carried analytes from the pressurized container to a Fluigent two-way switch placed before the prepolarizing Halbach array. Following the switch, tubing went into the Halbach array and was wound in several loops before exiting. The total volume of fluid in the Halbach was ${\sim} 0.1~{\rm mL}$. From the Halbach, ${\sim}35~{\rm \upmu L}$ of tubing ran to the microfluidic chip housing the diamond. Additional PEEK tubing was connected from the outlet of the chip, passed through a second fluidic switch (synchronized with the first switch) and exited into an exhaust container. 
 
 For the NMR coil magnetometer, a similar microfluidic path was built with the exception that no switch  was installed for stop-flow. Water flowed continuously through this line throughout all experiments. 
 
The timing of the fluidic switches was synchronized with the RF pulses and NV detection scheme. First, the switches were turned to the flow position for $1.25~{\rm s}$ to replenish the NV NMR detection region with freshly polarized analyte. The switches were then turned off, followed by a variable delay to allow for settling of the fluid at the diamond interface. Thereafter, the RF pulse sequence was applied, followed by a minimum detection period of 1.25 s for NMR detection. At the end of the NMR detection sequence, a final RF pulse was applied for the feedback NMR coil magnetometer (NV detection is OFF at this stage), and the switches were activated for reflow. The entire sequence was repeated in a loop to enable signal averaging.

\section{Adiabaticity considerations}
\label{sec:ad}
For optimum polarization retention, it is important to ensure that the spins remain aligned with the external magnetic field while transitioning from the prepolarization region to the detection region. This adiabaticity condition holds if the rate of change in the magnetic field direction is much smaller than the nuclear spin angular frequency \cite{Tayler2017}. To estimate if this condition holds for our experiments, we compare the proton spin angular frequency at the smallest field experienced during transit (${\sim}0.3~{\rm mT}$, corresponding to a spin angular frequency of ${\sim}8{\times}10^4~{\rm rad/s}$) to the rate of change of field angle experienced by the fastest moving spins in the microfluidic system.

The narrowest tubing (corresponding to fastest flow) used in this experiment is $R=90~{\rm \upmu m}$ radius PEEK tubing and the fastest volumetric flowrates used is $Q=50~{\rm \upmu L/s}=50~{\rm mm^3/s}$. Assuming laminar flow, the peak velocity at the center of the tube is:
\begin{equation}
    v_{max}=\frac{2 Q}{\pi R^2}\approx4\times10^3~{\rm mm/s}.
\end{equation}
The most abrupt change in field angle occurs in our setup immediately after the Halbach array, where the field angle changes by ${\sim}\pi/2$ radians over ${\sim}40~{\rm mm}$. The fluid thus experiences a rate of change in field angle of approximately $\pi/2~{\rm rad.}\frac{4{\times}10^3~{\rm mm/s}}{40~{\rm mm}}=150~{\rm rad./s}$. This angular rate is more than two orders of magnitude smaller than the spin precession angular rate, meaning the spins follow the field adiabatically.

\section{Optimization of Flow Rates}
\begin{figure}[hb]
    \includegraphics[width=0.95\columnwidth]{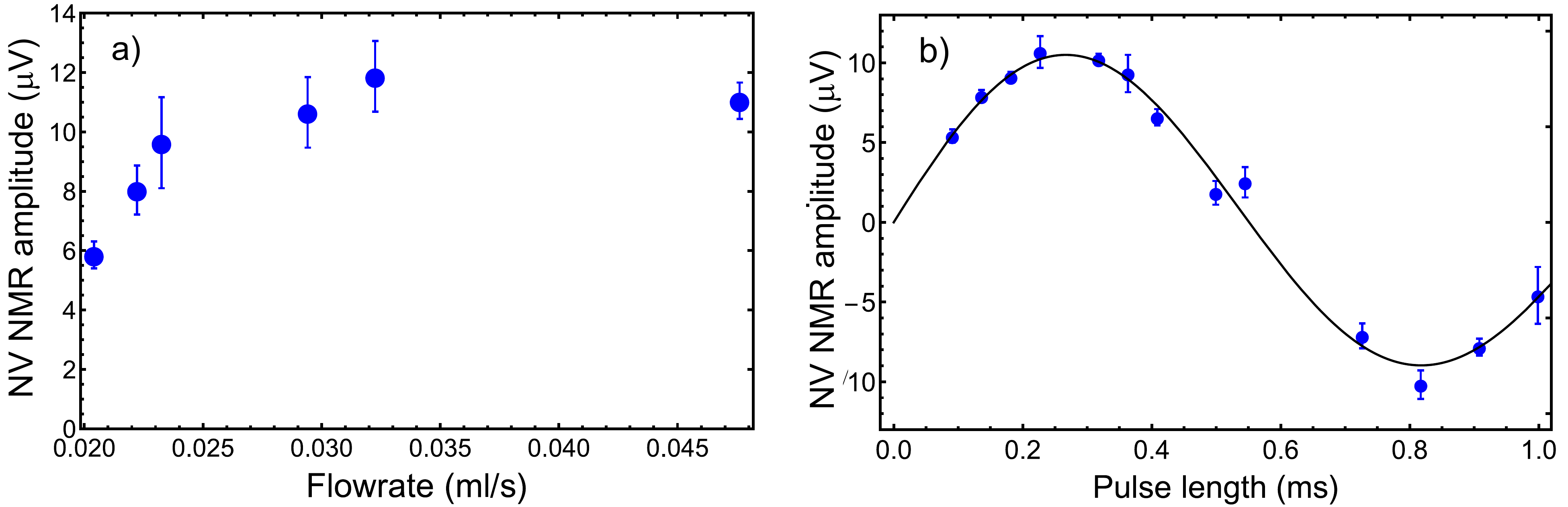}
    \caption{a) Dependence of water NV NMR amplitude on flowrate. The experiment was run by steadily increasing the pressure of the container and allowing for a reflow time of 3.5 s. The long flow time was chosen to exceed $T_1$ (${\approx}2~{\rm s}$ for water) to decouple the influence of $T_1$ and flow. b) Water NV NMR amplitude as a function of applied RF pulse length. The sinusodial nature of the curve indicates that the water proton spins are coherently driven by the RF pulses.}
    \label{fig:nuc_rabi}
\end{figure}
The analyte flow rate was optimized to efficiently transfer nuclear spin polarization from the Halbach array to the NV detection region. The ideal flow conditions ensure that there is sufficient ``residence'' time in the Halbach array, to fully polarize the sample, followed by fast delivery to the detection region before significant $T_1$ relaxation occurs. We measured the water NV NMR amplitude as a function of the flow-rate to find the optimal values, Fig.~\ref{fig:nuc_rabi}(a). Flow rates in the $25\mbox{--}50~{\rm \upmu L}$ range led to optimal polarization transfer. Flow rates were kept in this range for the reported experiments. 

To calibrate the nuclear $\pi/2$ pulse length, we measured the water NV NMR amplitude as a function of the applied RF pulse length, Fig.~\ref{fig:nuc_rabi}(b). The well-behaved sinusodial dependence is a sign that analyte was fully stopped within the excitation loop volume before detection. In continuous flow (data not shown) we did not see this behavior; instead a slowly building signal with a very slow decay was observed indicating that new sample was entering the excitation and detection volumes on the timescale of the experiment. No clear $\pi/2$ pulse length could be inferred. Thus using stop-flow was particularly important for the 2D NMR experiments, where well-defined nuclear pulse sequences are required.

\section{Magnetic field calibration}
\label{sec:magcal}
In Fig.~\ref{fig:pulse_seq_and_SNR}(b) of the main text, an AC test magnetic field was applied to calibrate the NV NMR response. The amplitude of this test field was calibrated using two complementary methods. In the first method, a DC current was applied to the test loop and the shift of the NV ODMR peaks was used to infer the conversion between current and the projection of the AC field amplitude along the NV axis, $B_{\rm AC}$. In the second procedure, $B_{\rm AC}$ was calibrated by determining the AC current amplitude needed to maximize the oscillation amplitude of the NV NMR signal, i.e. where the phase accumulation during a single XY8-5 sequence is equal to $\pi/2$. 

The time-domain NV NMR fluorescence signal depends on the AC magnetic field amplitude, $B_{\rm AC}$ as:
\begin{equation}
    F(t)=F_0[1+C\sin{\left(4B_{\rm AC}\gamma_{NV}\tau_{\rm tot}\right)}\cos{\left(2\pi(f-f_{\rm alias}) t +\phi_0\right)}],
\end{equation}
where $F_0$ is the mean fluorescence intensity, $C\approx0.01$ is the maximum fluorescence contrast of the XY8-5 sequence, $f$ is the frequency of the AC magnetic field, $f_{\rm alias}$ is the aliasing frequency of the readout sequence, $\tau_{\rm tot}=24.16~{\rm \upmu s}$ is the total phase accumulation time during a single XY8-5 sequence, and $\phi_0$ is the phase of the applied oscillating magnetic field at the start of the acquisition sequence. We triggered the test field so that $\phi_0$ is the same for each experiment, as is the case for detection of the AC nuclear field.

Evidently the maximum oscillation amplitude of $F(t)$ occurs for values of $B_{\rm AC}$ that satisfy $\sin{\left(4B_{\rm AC}\gamma_{NV}\tau_{\rm tot}\right)}=1$. This occurs when:
\begin{equation}
    B_{\rm AC, max} = \pi/(8\gamma_{NV}\tau_{\rm tot}).
\end{equation}

\begin{figure}[hb]
    \centering
    \includegraphics[width=0.5\textwidth]{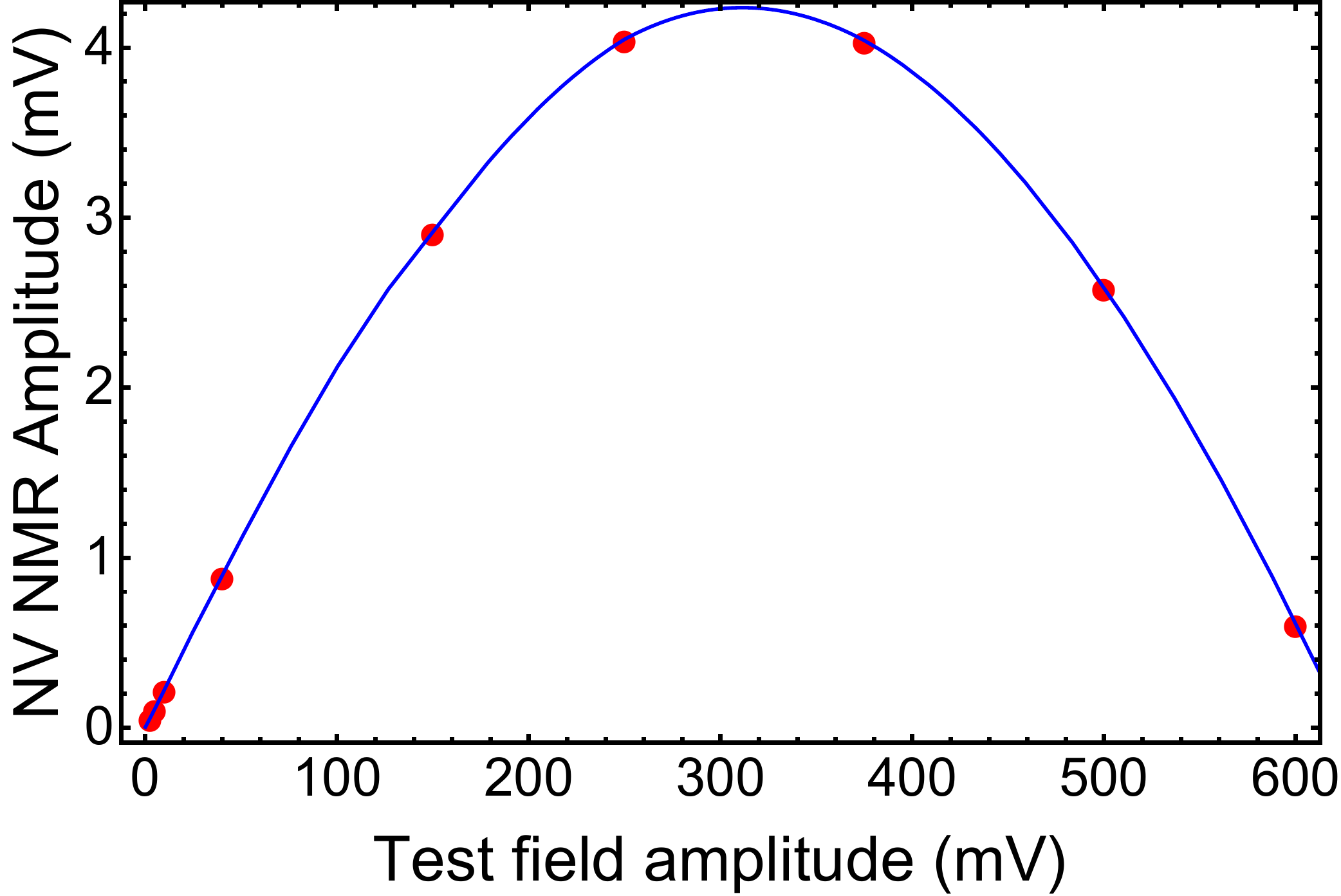}
    \caption{Saturation curve of the NV NMR spectrometer used for calibrating the magnetic field of AC test signals.}
    \label{fig:sens_calib}
\end{figure}

To determine the conversion of test field current amplitude and magnetic field, we acquired NV NMR sequences for different values of the voltage applied to the test loop. We determined the fluorescence oscillation signal amplitude by fitting the Fourier transform of the NV signal. Fig.~\ref{fig:sens_calib} shows a plot of the NV signal amplitude as a function of test field amplitude. From this measurement we were able to calibrate the magnetic field value of the applied test signal used in Fig.~\ref{fig:pulse_seq_and_SNR}(b) of the main text.

\section{Concentration sensitivity}
\label{sec:concsen}
In Fig.~\ref{fig:pulse_seq_and_SNR}(b) of the main text, the data were acquired for $1.25$ s while the experiment was repeated every $2.5$ s for a total of 60 scans. To calculate the concentration sensitivity, we used only the first $0.087~{\rm s}$ of NV NMR data, so the effective acquisition times was $60{\times}0.085~{\rm s}=5.2~{\rm s}$. The final concentration sensitivity of $27~{\rm M~s^{1/2}}$ was calculated using only this effective acquisition time, while neglecting dead time from the remainder of the pulse sequence and flow time. This estimate of concentration sensitivity is valid if one assumes that the transit time between prepolarization and detection phases can be neglected and the fluid can be flowed and stopped instantaneously. We anticipate that the transit times could be several orders of magnitude faster than we used here by optimizing the experimental geometry. Fluidic switches with 20 ms rise/fall time are commercially available, so this estimate is within reach for an optimized experiment.

\begin{figure}[ht]
    \centering
    \includegraphics[width=0.5\textwidth]{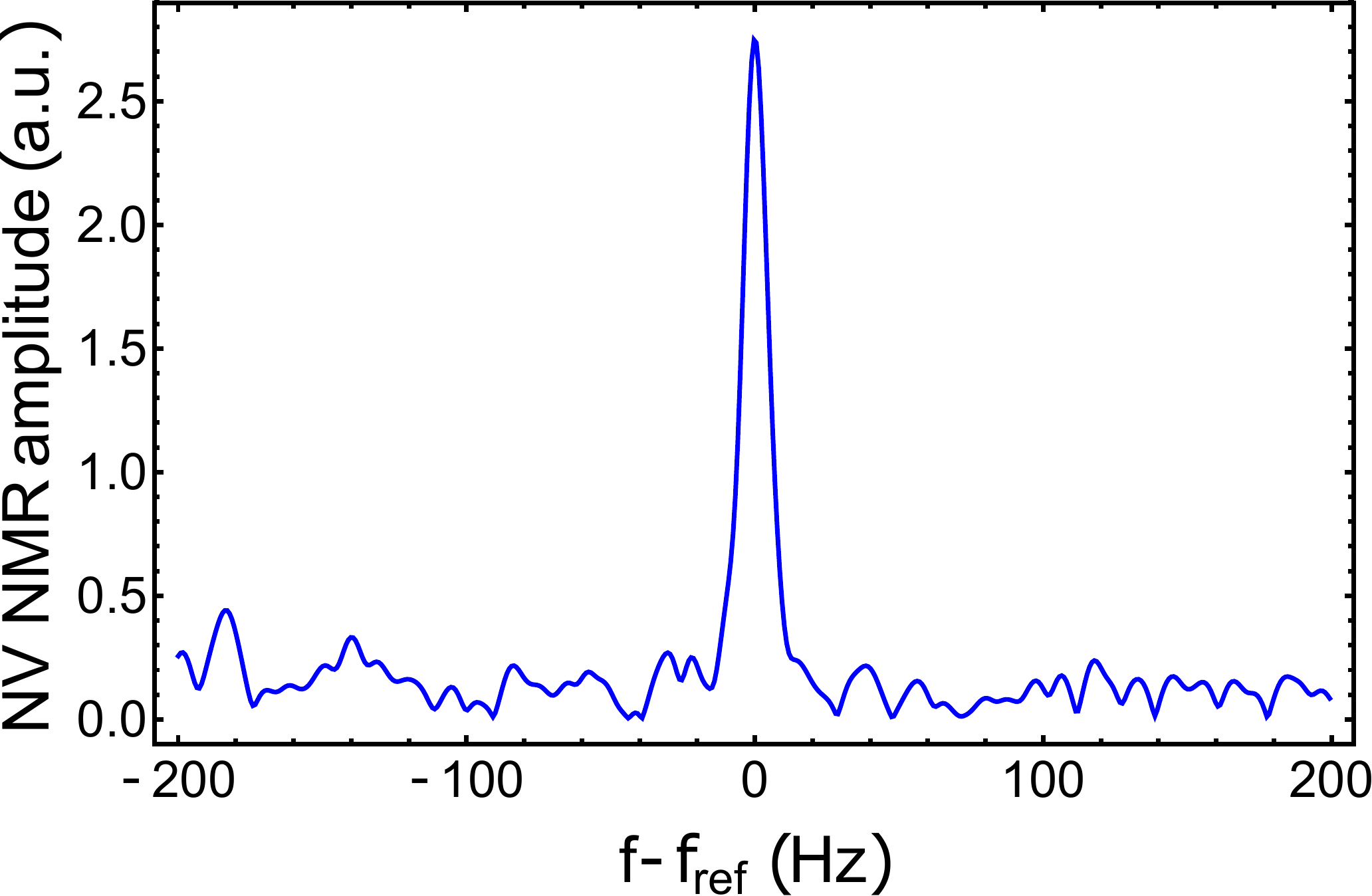}
    \caption{NV NMR spectrum of water. A Lorentz-to-Gauss windowing function is applied with $\alpha = 16~{\rm s^{-1}}$ and $\beta = 13.5~{\rm s^{-1}}$. The spectrum represents an average of 60 scans, with a scan repetition period of $2.5~{\rm s}$, corresponding to a total experimental acquisition of $150~{\rm s}$. The computed SNR is $88$, corresponding to a concentration sensitivity of $45~{\rm M~s^{1/2}}$.}
    \label{fig:real_snr}
\end{figure}

To calculate the more conservative concentration sensitivity, incorporating all experimental dead times, we use the full NV NMR signal and apply a windowing function to optimize SNR. Specifically, we apply a Lorentz-to-Gauss windowing function of the form:
\begin{equation}
\begin{split}
& W(t) = e^{\alpha t}e^{-(\beta t)^2},\\
&\alpha \geq 0, \beta \geq 0,
\label{eq:windowing}
\end{split}
\end{equation}
to the time domain data and compute the Fourier Transform, Fig.~\ref{fig:real_snr}. The signal is then defined as the discrete sum of the Fourier signal in a [$-7~{\rm Hz},+7~{\rm Hz}$] interval around the central maximum. The noise is defined as the standard deviation of the signal-free part of the spectrum multiplied by the square root of the number of points in the signal window. Using this method we find a SNR of 88 for the full $60\times2.5~{\rm s}=150~{\rm s}$ duration of the experiment. This corresponds to a SNR of 7.2 for 1 second integration or a concentration sensitivity of $45~{\rm M~s^{1/2}}$. This conservative estimate is less than a factor of two worse than the optimal estimate ($27~{\rm M~s^{1/2}}$).

\section{NMR field amplitudes and effective sensing volume}
\label{sec:nmrsig}
A magnetostatic model was developed to estimate the water nuclear AC field amplitude and effective proton detection volume. The illuminated NV sensor volume was modeled as a cylinder with a $20~{\rm \upmu m}$ diameter and $35~{\rm \upmu m}$ height defined by the thickness of the diamond and the laser beam area. The water proton volume was modeled as a hemisphere just above the sensor. Curie's law was used to estimate the magnetization:
\begin{equation}
    M = \frac{\rho \mu^2 B_H}{k_B T},
\end{equation}
where $\rho=6.7{\times}10^{28}~{\rm m^{-3}}$ is the proton spin density, $\mu$ is the proton magnetic moment, $k_B$ is the Boltzmann constant, $T=300~{\rm K}$ is the temperature, and $B_H=1.5~{\rm T}$ is the polarizing magnetic field. The model outputs the component of the nuclear AC magnetic field along the NV axis, integrated over the NV sensor volume.

\begin{figure}[hb]
    \centering
    \includegraphics[width=0.45\textwidth]{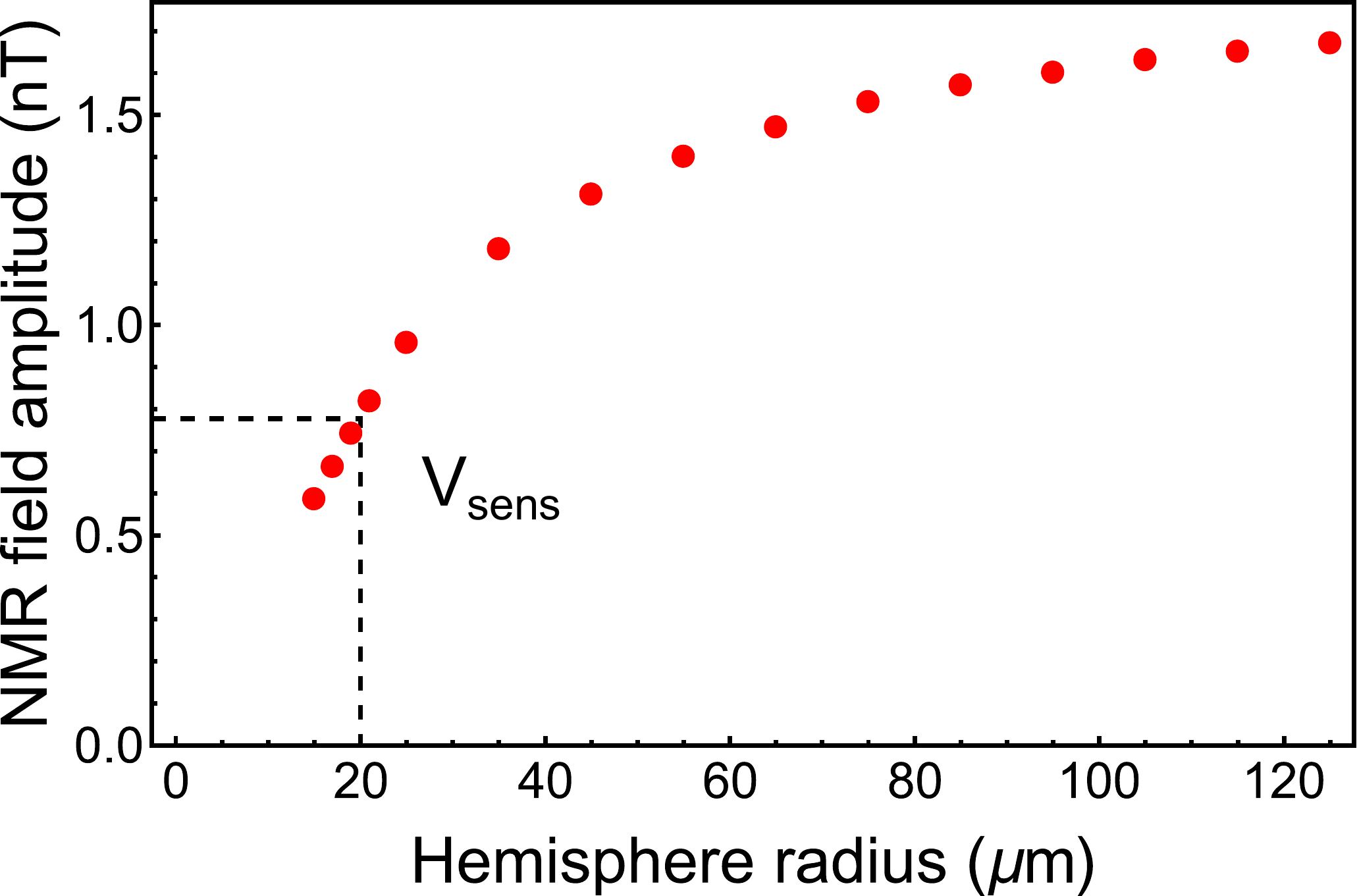}
    \caption{Nuclear AC magnetic field amplitude (integrated across the sensor volume) as a function of proton volume. The effective sensing volume $V_{sens}\approx20~{\rm pL}$ is defined as the volume where the nuclear AC field amplitude is equal to half of that in the case of very large volumes.}
    \label{fig:SI_sensing}
\end{figure}

The dependence of nuclear AC field amplitude on proton volume is shown in Fig.~\ref{fig:SI_sensing}. We define the effective proton detection volume as the volume of protons that generate a nuclear field amplitude (integrated over the NV sensor volume) equal to half of the asymptotic limit for large proton volumes $\gtrsim10~{\rm nL}$. By this definition, the proton detection volume is ${\sim}20~{\rm pL}$ and the component of the nuclear AC magnetic field amplitude along the NV axis is $1.6~{\rm nT}$ for full nuclear spin polarization ($5.1{\times}10^{-6}$). In Fig.~\ref{fig:pulse_seq_and_SNR}(b) of the main text, we observe a nuclear AC field projection amplitude of $1.21~{\rm nT}$. This is $77\%$ of the maximum field anticipated from magnetostatic modeling, indicating that an effective polarization of $3.9{\times}10^{-6}$ is detected.

\section{Analytical calculation for heteronuclear COSY}
\label{sec:2dsim}

To gain intuition for the features present in the heteronuclear COSY spectra in Fig.~\ref{fig:2D_dfb}(e,f), we performed an analytical calculation on a simplified two-spin system with spin angular momenta $I_1=1/2$ and $I_2=1/2$ \cite{halse2009multi}. Note that, in our case, $I_1$ and $I_2$ correspond to nuclei of different isotopes, e.g. $^1{\rm H}$ and $^{19}{\rm F}$. A magnetic field is applied along the $z$ direction, and the spin Hamiltonian is:
\begin{equation}
\mathcal{H}/\hbar=I_{z1}\omega_1+I_{z2}\omega_2+2\pi J_{12}\vec{I_1}\cdot\vec{I_2},
\end{equation}
where $\omega_1$ is the spin precession angular frequency of spin 1 and $\omega_2$ is that of spin 2. As $\vert \omega_1 -\omega_2 \vert \gg 2\pi J_{12}$ we can apply the secular approximation:
\begin{equation}
\mathcal{H}/\hbar =I_{z1}\omega_1+I_{z2}\omega_2+2\pi J_{12}I_{z1}I_{z2}.
\end{equation}
We assume the spins are initially polarized along the magnetic field, resulting in an initial state: 
\begin{equation}
I_{z1}+I_{z2}.
\end{equation} 

We now consider the heteronuclear COSY pulse sequence beginning with a $(\pi/2)_x$ pulse acting on spin 1, followed by an evolution time $t_1$, and a second $(\pi/2)_x$ pulse acting on spin 2. Immediately following the first $(\pi/2)_x$ pulse acting on spin 1, the state is:
\begin{equation}
I_{y1}+I_{z2}.
\end{equation}
The evolution under the $J$-coupling and Zeeman terms can be applied sequentially, as both operators commute. The state evolves under $J$-coupling as:
\begin{equation}
\cos{\left(\pi J t_1\right)}I_{y1}-2\sin{\left(\pi J t_1\right)} I_{x1}I_{z2} + I_{z2}.
\end{equation}
This state evolves in the external field as:
\begin{equation}
\begin{split}
I_{z2}+\cos{\left(\pi J t_1\right)}\left[\cos{\left(\omega_1 t_1\right)}I_{y1}+\sin{\left(\omega_1 t_1\right)}I_{x1}\right] \\ -
2\sin{\left(\pi J t_1\right)}\left[\cos{\left(\omega_1 t_1\right)}I_{x1}I_{z2}-\sin{\left(\omega_1 t_1\right)}I_{y1}I_{z2}\right].
\end{split}
\end{equation}
Next, we apply the last $(\pi/2)_x$ pulse on spin 2 and the state becomes:
\begin{equation}
\begin{split}
        \cancel{I_{y2}}+\cos{\left(\pi J t_1\right)}\left[\cos{\left(\omega_1 t_1\right)}I_{y1}+\sin{\left(\omega_1 t_1\right)}I_{x1}\right]  \\ -
  2\sin{\left(\pi J t_1\right)}\left[      \cancel{\cos{\left(\omega_1 t_1\right)}I_{x1}I_{y2}}-\cancel{\sin{\left(\omega_1 t_1\right)}I_{y1}I_{y2}}\right],
\end{split}
\end{equation}
Here the slashes over terms show the terms that we can drop because they have no time dependence during the evolution period $t_1$ or they have a coherence order other than $p=\pm 1$.

To understand how the state evolves during the second evolution period we again apply the $J$-coupling and Zeeman evolution operators on the observable states.
Evolution under $J$-coupling leaves the state as:
\begin{equation}
\begin{split}
    \cos{\left(\pi J t_1\right)}\left[\cos{\left(\omega_1 t_1\right)}\left\{\cos{\left(2\pi J t_2\right)}I_{y1}-\cancel{2\sin{\left(2\pi J t_2\right)}I_{x1}I_{z2}}\right\} \right. \\ + \left.
    \sin{\left(\omega_1 t_1\right)}\left\{\cos{\left(2\pi J t_2\right)}I_{x1}+\cancel{2\sin{\left(2\pi J t_2\right)}I_{y1}I_{z2}}\right\}\right].
\end{split}
\end{equation}
Here we have eliminated the antiphase terms which do not contribute meaningfully to the spectrum. Finally, evolution under the Zeeman interaction leaves the state as:
\begin{equation}
\begin{split}
\cos{\left(\pi J t_1\right)}\cos{\left(\pi J t_2\right)}\cos{\left(\omega_1 t_1\right)}\left[\cos{\left(\omega_1 t_2\right)}I_{y1}+\sin{\left(\omega_1 t_2\right)}I_{x1}\right] \\ +
\cos{\left(\pi J t_1\right)}\cos{\left(\pi J t_2\right)}\sin{\left(\omega_1 t_1\right)}\left[\cos{\left(\omega_1 t_2\right)}I_{x1}-\sin{\left(\omega_1 t_2\right)}I_{y1}\right].
\end{split}
\end{equation}
Gathering terms by spin operators, we obtain:
\begin{equation}
\begin{split}
\cos{\left(\pi J t_1\right)}\cos{\left(\pi J t_2\right)}\left\{\left(\cos{\left(\omega_1 t_1\right)}\cos{\left(\omega_1 t_2\right)}-\sin{\left(\omega_1 t_1\right)}\sin{\left(\omega_1 t_2\right)}\right)I_{y1} \right. \\ \left.
+\left(\cos{\left(\omega_1 t_1\right)}\sin{\left(\omega_1 t_2\right)}+\sin{\left(\omega_1 t_1\right)}\cos{\left(\omega_1 t_2\right)}\right)I_{x1}
\right\}
\end{split}
\end{equation}
Evidently all terms are modulated by $ \cos{\left(\pi J t_1\right)}\cos{\left(\pi J t_2\right)}$. This is what leads to the off diagonal terms in the heteronuclear COSY experiments of Fig.~\ref{fig:2D_dfb}(e,f).

\section{2D homonuclear COSY of trimethyl phosphate}
\label{sec:tmp2d}
\begin{figure}[h]
    \centering
    \includegraphics[width=0.4\textwidth]{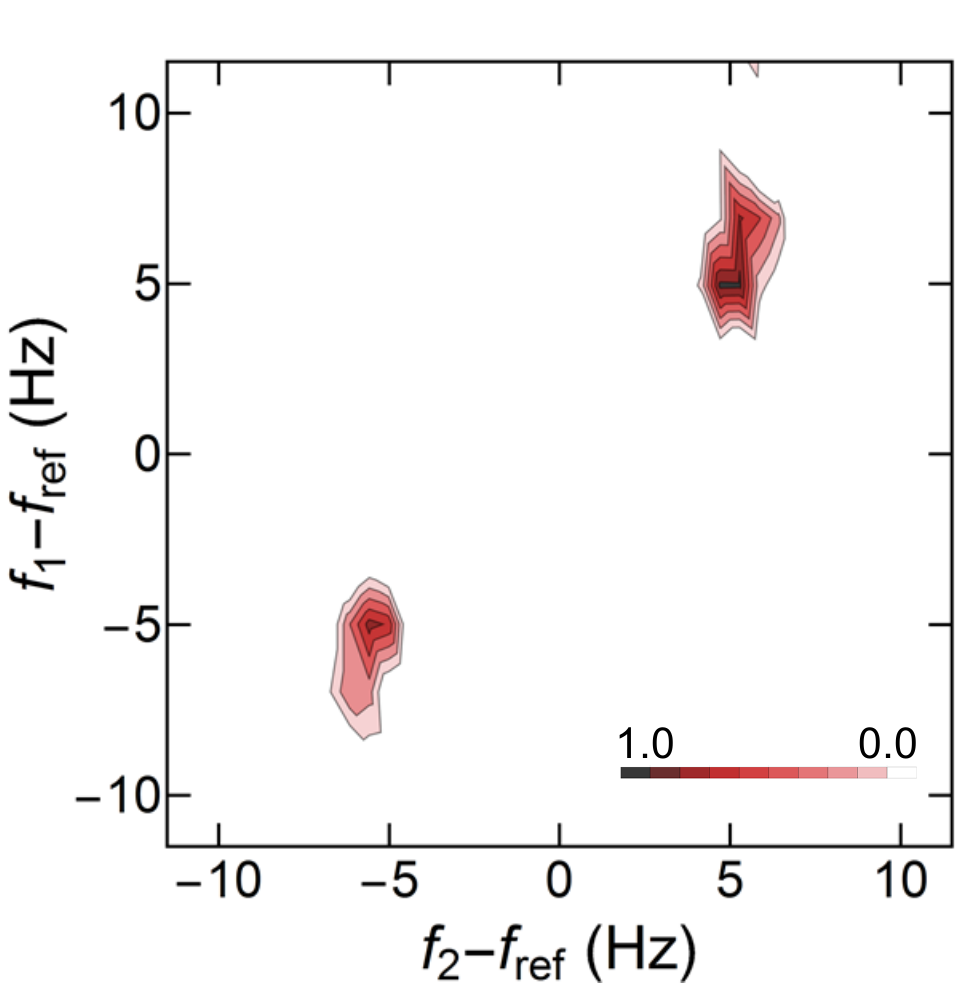}
    \caption{Experimental homonuclear COSY spectrum of trimethyl phosphate. The normalized absolute value of the 2D Fourier Transform is plotted.}
    \label{fig:TMP_2D}
\end{figure}
A 2D homonuclear COSY spectra was also obtained for trimethyl-phosphate, Fig.~\ref{fig:TMP_2D}.  We used 16 values of $t_1$ in 0.021 s increments up to 0.336 s. The total acquisition time was $12~$hours. The presence of two diagonal peaks indicates that magnetization is modulated at the $J$-coupling frequency during the $t_1$ evolution, but no magnetization is transferred between nuclei. Like for DFB, this is expected since there is no difference in chemical shift between protons.

\section{SPINACH simulations and windowing functions for 2D NMR}
The 2D NMR simulations were performed using the SPINACH simulation package \cite{hogben2011spinach}. The relaxation rate was set to 0 Hz. The $J$-couplings used in the simulation are listed in table \ref{tab:j_couplings}. The $o, m$ and $p$ suffixes indicate whether the two nuclei are in an ortho-, meta- or para- configuration relative to one another. Proton chemical shifts were assumed to be equal to zero. 
\begin{table}[h]
    \centering
    \begin{tabular}{|l|l|}
    \hline
        \textbf{Type of coupling} & \textbf{Coupling strength}\\ \hline
       J$^o[ ^{19} $F,$^{1}$H$]$  &  7.6 Hz  \\ \hline
         J$^m[ ^{19}$F,$^{1}$H$]$ & 4.6 Hz \\ \hline
         J$^o[^{1}$H,$^{1}$H$]$ & 8 Hz \\ \hline
         J$^m[^{1}$H,$^{1}$H$]$ & 2 Hz \\ \hline
         J$^p[^{1}$H,$^{1}$H$]$ & 0 Hz \\ \hline
         J$^p[^{19}$F,$^{19}$F$]$ & 12 Hz \\ \hline
    \end{tabular}
    \caption{Values of the different $J$-couplings in a DFB molecule used in the simulation \cite{paterson1964nmr}.}
    \label{tab:j_couplings}
\end{table}

Lorentz-to-Gauss transformations were applied to both the experimental and simulated DFB COSY spectra using the windowing function:
\begin{equation}
\begin{split}
& W(t_1, t_2) = e^{\alpha_1 t_1}e^{-(\beta_1 t_1)^2}e^{\alpha_2 t_2}e^{-(\beta_2 t_2)^2},\\
&\alpha_1 \geq 0, \beta_1 \geq 0, \alpha_2 \geq 0, \beta_2 \geq 0.   
\end{split}
\end{equation} The same $\alpha_{1,2}$ and $\beta_{1,2}$ parameters were used for the experimental and simulated spectra. For the heteronuclear dataset, we used $\alpha_{1,2} = 1.2~{\rm s^{-1}}$ and $\beta_{1,2} = 4~{\rm s^{-1}}$. This windowing process contributed ${\sim}3~{\rm Hz}$ to the width of the NMR lines in both the $f_1$ and $f_2$ dimensions. In the homonuclear dataset the parameters were $\alpha_1 = 3~{\rm s^{-1}}$, $\alpha_2 = 1.6~{\rm s^{-1}}$, $\beta_1 = 2.8~{\rm s^{-1}}$, and $\beta_2 = 2.5~{\rm s^{-1}}$, which contributed ${\sim}2.2~{\rm Hz}$ to the width of the NMR lines in both $f_1$ and $f_2$ dimensions.
\label{sec:spinach}

\end{document}